\documentstyle[prl,aps,floats]{revtex}
\input{psfig.tex}

\newcommand{\beq}{\begin{equation}}
\newcommand{\eeq}{\end{equation}}
\newcommand{\eea}{\end{eqnarray}}
\newcommand{\bea}{\begin{eqnarray}}

\begin{document}
%DAMTP-2001-94
\draft
\twocolumn[\hsize\textwidth\columnwidth\hsize\csname@twocolumnfalse\endcsname 
\title{Constraining dark energy with Sunyaev-Zel'dovich cluster surveys}
\author{Jochen Weller$^1$, Richard A. Battye$^2$, R\"udiger Kneissl$^3$}
\address{$^1$Department of Applied Mathematics and Theoretical Physics, Centre for Mathematical Sciences,\\ University of Cambridge, Wilberforce Road,
Cambridge CB3 0WA, U.K.\\$^2$Jodrell Bank Observatory, University of Manchester,  Macclesfield, Cheshire SK11 9DL, U.K.
\\$^3$Astrophysics Group, Cavendish Laboratory, University of Cambridge, Madingley Road, 
Cambridge CB3 0HE, U.K.}
\maketitle
\begin{abstract}

We discuss the prospects of constraining the properties of a dark
energy component, with particular reference to a
time varying equation of state, using future cluster surveys selected
by their Sunyaev-Zel'dovich effect.
We compute the number of clusters expected for a
given set of cosmological parameters and propogate the errors expected
from a variety of surveys. In the short term they will 
constrain dark energy  in conjunction with future
observations of type Ia supernovae, but may in time do so in
their own right. 

\end{abstract}
]
Recent observations of type Ia supernovae (SNe) have motivated the
search for a ubiquitous energy density component, known as dark
energy~\cite{sn}. The defining properties of this energy are that it
has negative pressure and does not
cluster into galaxies in the same way as dark matter, remaining
homogeneous on all but the largest scales. The standard form is the
cosmological constant 
($\Lambda$), although other possibilities exist including a slowly
rolling scalar field~\cite{Q}, known as Quintessence.

The quantification of the properties of this dark energy is now a
major part of many observational programs. One proposal is a
satellite, known as SNAP (SuperNova Acceleration Probe) \cite{SNAP} which should find around 1800 SNe out to $z\approx 1.7$. This will  certainly
constrain the properties of dark energy~\cite{Huterer:99,AW}, but without prior
information on the matter density, $\Omega_{\rm m}$, this will have
very little to say about the time evolution of the equation of state
parameter $w_{\phi}=p_{\phi}/\rho_{\phi}$, crucial for distinguishing between the various dark energy
models~\cite{AW}. In this {\it Letter}, we discuss another approach
using future cluster surveys selected using the  Sunyaev-Zel'dovich (SZ)
effect. We will show that, dependent on the
angular coverage ($\Delta\Omega$),
frequency ($\nu$) and flux limit
($S_{\rm lim}$), such a survey  may provide complementary
information to SNe observations, or accurately constrain the properties of the
dark energy in its own right.

Observations of clusters via the SZ effect~\cite{Sunyaev:72} (see
ref.~\cite{birk} for a recent review) exploit the fact that the
cosmic microwave background (CMB) radiation is rescattered by hot
intracluster gas.
Since Compton scattering conserves the overall
number of photons, the radiation gains energy by redistributing them
from lower to higher frequencies. If one observes them in the
Rayleigh-Jeans region of the spectrum, the flux of observed photons
decreases compared to the unscattered CMB radiation. 
The total flux depends on the gas mass and mean temperature 
of the cluster, but is independent of their distributions.
Moreover, the number density of such clusters evolves 
with redshift under the action of gravity 
making it an ideal probe of cosmology~\cite{ECF}.

The first step is to compute the distribution of clusters
which will be observed by a particular survey for a given set of cosmological 
parameters.
We choose to consider the redshift distribution of clusters with mass
larger than $M_{\rm lim}\propto \{S_{\rm
lim}d_{\rm A}^2/[\nu^2(1+z)]\}^{3/5}H(z)^{-2/5}$ which is given by
\beq
	\frac{dN}{dz} =
{\Delta\Omega}\frac{dV}{dzd\Omega}(z)\int\limits_{M_{\rm 
lim}(z)}^\infty \frac{dn}{dM}\,dM\, ,
\eeq
with  $\frac{dn}{dM}dM$ the comoving density of clusters with mass between $M$
and $M+dM$, $\frac{dV}{dzd\Omega}$ the volume element, $H(z)$ the Hubble parameter, and $d_{\rm A}$
the angular diameter distance. The
distribution $\frac{dN}{dS}$ could
also constrain cosmological parameters and may be powerful if there is
sparse redshift information available. However, we do not expect it to be
very sensitive to the equation of state of dark energy since the 
crucial redshift dependence is integrated out. 

The limiting mass $M_{\rm lim}$
of the survey can be  directly related to the total limiting flux $S_{\rm lim}$ of the SZ
survey by the virial theorem and the SZ flux~\cite{Viana:96,Haiman:01,Weller:01b}. We assume that
the geometry of the universe is flat and that the late time dynamics
is dominated by a matter component with density $\Omega_{\rm m}$ and a
dark energy component with $\Omega_\phi = 1-\Omega_{\rm m}$. Since
there are a wide range of dark energy models discussed in the literature
(see refs.~\cite{Q,AW,BM} and references therein) which all have potentially different late time behaviour we choose to parametrize the equation of 
state by its late time evolution $w_\phi=w_0+w_1z$. The comoving
number density is taken from a series of N-body simulations~\cite{Jenkins:01}, which
yields results similar to using the  Press-Schechter formalism
~\cite{Press:74}, but predicts more massive and less
`typical' clusters, as observed in the simulations~\cite{Pierpaoli:01}.
%\bea
%\frac{dn}{dM}\left(z,M\right) &=& - 0.315
%\frac{\rho_{\rm m}(0)}{M}\frac{d\sigma_M}{dM}\frac{1}{\sigma_M}\times
%\nonumber \\
%& &\exp\left\{-|0.61-\log\left[D(z)\sigma_M\right]|^{3.8}\right\}\, ,
%\label{eqn:conum}
%\eea
The linear growth factor is computed for a given cosmology by
solving the ODE for the linear perturbations~\cite{Viana:96}
numerically and non-linear evolution is taken into account via the
spherical collapse model. 

\begin{figure}[!h]
\setlength{\unitlength}{1cm}
\centerline{\psfig{file=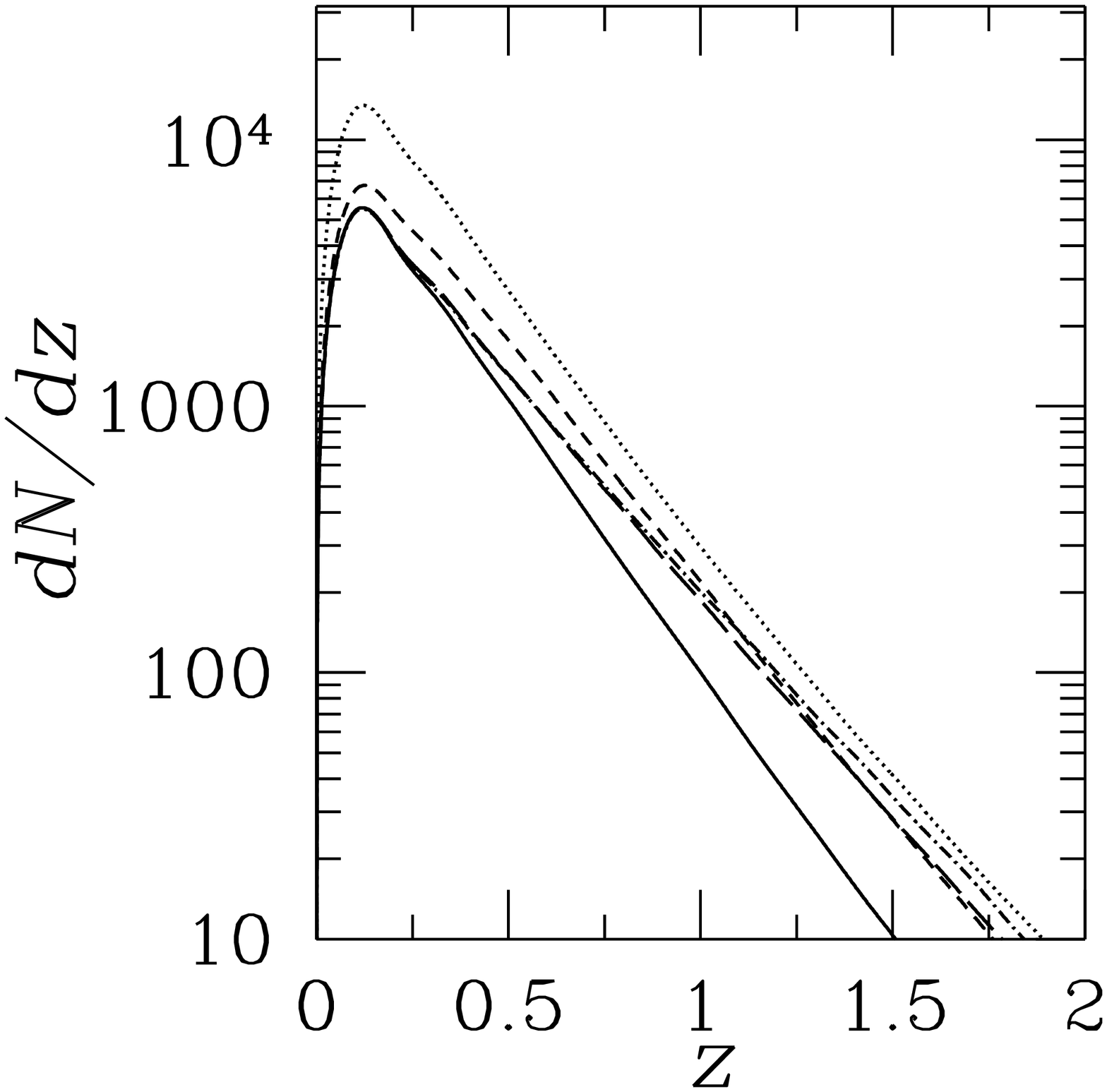,width=4.5cm,height=4.5cm}\psfig{file=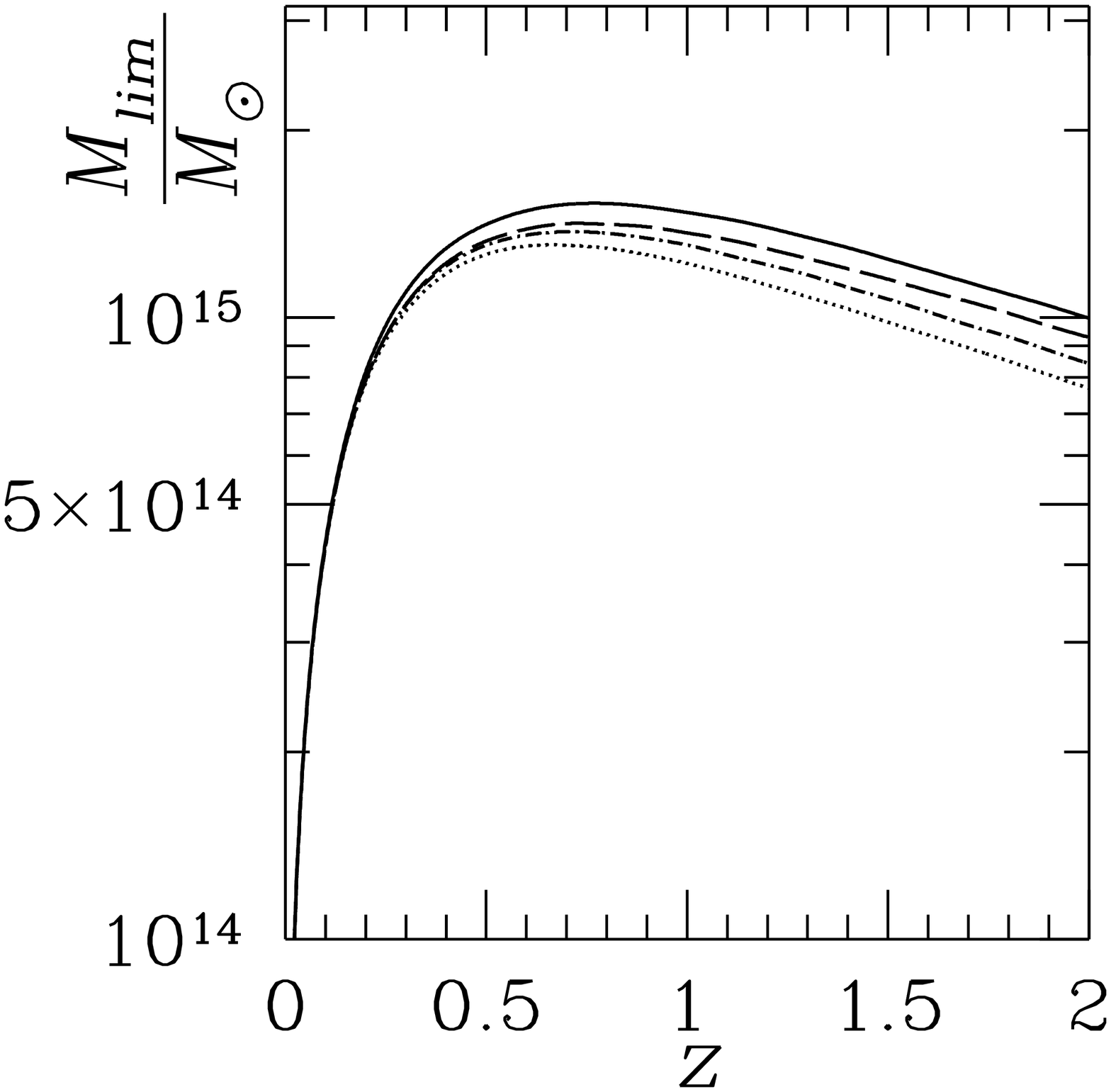,width=4.5cm,height=4.5cm}}
\vskip 5pt
\caption{In the left panel we show the cosmology dependence of the
redshift evolution of the number of clusters and in the right panel
the mass threshold. We vary $\Omega_{\rm m}$, $\sigma_8$, $w_0$ and
$w_1$ as explained in the text. The results correspond to the
experimental setup (II) with $10^4\,{\rm deg}^2$ sky coverage.}
\label{fig:dNdz}
\end{figure}

In Fig.~\ref{fig:dNdz} we illustrate the dependence of the
redshift distribution of SZ clusters and the limiting mass on cosmology.
The solid line is a model with $\Omega_{\rm m}=0.3$, the Hubble constant $H_0 = 65 {\rm km}\,{\rm sec}^{-1}{\rm Mpc}^{-1}$, $\sigma_8=0.925$, $w_0=-1$, 
$w_1 = 0$ and 
spectral index of density fluctuations $n=1$. The dotted line has
$\Omega_{\rm m}=0.5$, the dashed line $\sigma_8=0.975$, the long dashed line
$w_0=-0.8$ and the dot-dashed line $w_0 = -0.8$ and $w_1=0.3$.  The
dependence on $n$ is very weak~\cite{Haiman:01} and we therefore fix $n=1$.
 We will consider the possible dependence of the number
density on the parameters $\Theta=(H_0,\,\sigma_8,\,\Omega_{\rm
m},\,w_0,\,w_1)$ in the subsequent analysis.
From Fig.~\ref{fig:dNdz} we see that
$dN/dz$ is strongly dependent on $\Omega_{\rm m}$ and
$\sigma_8$, while the dependence on $w_0$ is still recognizable and
that on $w_1$ is relatively weak. 

We make the optimistic assumption that all the clusters found 
in the complete surveys can be 
located sufficiently well so as to determine their redshift out to
some critical value $z_{\rm max}$. Furthermore, we
will assume that this will be known within a precision of $\Delta
z=0.01$ which will allow us to use data bins of size $\Delta z$. This level of accuracy will only require the redshifts to be determined photometrically and will be possible using SDSS (Sloan Digital Sky Survey) and VISTA (Visible and Infrared Survey Telescope for Astronomy). We can then 
compare to theoretical models using  the  Cash C
statistic~\cite{Cash:79,Holder:01b} for the log-likelihood assuming that the errors are Poisson distributed. 

A number of surveys are expected which are designed to detect all 
clusters above some limiting mass $M_{\rm lim}(z)$. For the purposes of our discussion we will group
them into four categories whose observing strategies, approximate
$M_{\rm lim}$ and projected number of observed number of clusters in
a dark energy based cosmology are tabulated in Table I.
\begin{table}[t]
\centering
\begin{tabular}{|l | c | c | c | c |}
 &(I) & (II) & (III) & (IV) \\
\hline
$S_{\rm lim}$ & 0.1 & 5 & $\approx$36 & - \\
$\nu$ & 15 & 30 & $\approx$100 & - \\
$\Delta\Omega$ & 10 & $10^4$ & 20600 & 4000\\
$M_{\rm lim}$ & $1.5$ & $\approx 7.0$ & $\approx 6.0$ & 2.5 \\
$N_{\rm tot}$ & $\approx 90$ & $\approx 1970$ & $\approx 5200$ & $\approx 13600$ \\
\hline
$\delta H_0$ &$\pm\infty$ &$\pm 15$ &$\pm 15$ &$\scriptstyle{-10 / +5}$\\
$\delta\sigma_8$ &$\pm 0.075$ &$\pm 0.03$ & $\pm 0.02$ & $\pm 0.007$\\
$\delta\Omega_{\rm m}$ &$\scriptstyle{-0.07 / +0.10}$ &$\pm 0.05$ & $\pm 0.03$ & $\pm 0.02$ \\
$\delta w_0$ &$\scriptstyle{-\infty/ +0.30}$ &$\scriptstyle{-0.15/+0.29}$ & $\scriptstyle{-0.09 / 
+0.12}$ & $\scriptstyle{-0.04 / + 0.12}$\\
$\delta w_1$ &$\pm\infty$ & $\scriptstyle{-0.60/+0.14}$ & $\scriptstyle{- 0.46 / + 
0.10}$ & $\scriptstyle{-0.55 / +0.05}$\\ 
\hline 
$\delta H_0$ &$\scriptstyle{-15 / +5}$ &$\pm 5$ &$\pm 7$ & $\pm 5$\\
$\delta\Omega_{\rm m}$ &$\scriptstyle{-0.04 / +0.08}$ &$\pm 0.03$ &$\pm 0.02$ & $\pm 0.01$ \\
$\delta w_0$ &$\scriptstyle{-0.07 / + 0.28}$  &$\scriptstyle{-0.03 /
+0.14}$ & $\scriptstyle{-0.09 / +0.12}$ & $\pm 0.03$\\
$\delta w_1$ &$\scriptstyle{-\infty / +0.15}$ &$\scriptstyle{-0.47 /
+0.09}$ & $\scriptstyle{-0.42 / + 0.06}$ & $\pm 0.03$\\
\hline 
$\delta\Omega_{\rm m}$ &$\pm 0.04$ &$\pm 0.02$ &$\pm 0.01$ & $\pm 0.01$\\
$\delta w_0$ &$\scriptstyle{-0.07 / +0.23}$ &$\scriptstyle{-0.02 /
+0.10}$ & $\scriptstyle{-0.07 / + 0.04}$ & $\pm 0.03$\\
$\delta w_1$ &$\scriptstyle{-\infty / +0.15}$  &$\scriptstyle{-0.40 /
+0.04}$ & $\scriptstyle{-0.28 / + 0.07}$ & $\pm 0.03$\\
\end{tabular}
\vskip 0.5cm
\caption{The properties of the different classes of experiments, the number of
clusters one would expect to observe in a fiducial cosmology 
and the 1-$\sigma$ uncertainties on the parameters one would deduce
for the same cosmology if one (a) had no prior information,
(b) with fixed $\delta\sigma_8$ and (c) imposed both $\delta H_0=5$ and
fixed $\delta\sigma_8$ together. The units of
($H_{0}$,$\,S_{\rm lim}$,$\,\nu$,$\,\Delta\Omega$,$\,M_{\rm lim}$) are
(${\rm km}\,{\rm sec}^{-1}{\rm Mpc}^{-1}$,$\,$mJy,$\,$GHz,$\,{\rm
deg}^2$,$\,10^{14}h^{-1}M_\odot$). We used $\infty$ to denote cases where we
were unable to make a sensible constraint on a particular parameter.}
\end{table}
The first category (I) of deep and narrow surveys contains the interferometric arrays 
AMI~\cite{AMI} (Arc-minute Micro-Kelvin Imager), SZA 
~\cite{SZA} (SZ Array) and AMiBA~\cite{Lo:00} (Array
for Microwave Background Anisotropy). For AMI detailed simulations of the survey yield have been performed and radio source contamination has been considered. The second group (II) are
shallow and wide surveys of which  OCRA~\cite{OCRA} (One-Centimetre
Receiver Array) is an example. Here, we use the flux sensitivity for a single receiver from the proposed 
100 beam array, without taking into account the effects of atomospheric water vapour at the site. 
The third class (III), which are shallow but nearly all-sky,
correspond to what might be possible based on component separation
using the multi-frequency 
channels of the PLANCK surveyor. As an example of the sensitivity
we list the 100 GHz channel. The final category (IV), are deep and
wide surveys, such as a 1000 element bolometric array which may be mounted on a telescope at the south pole. 
In the last case, due to lack of precise figures we will use a
constant limiting mass~\cite{Holder:01b}.

The 1-$\sigma$ errors one would expect on $\delta\Theta$ are listed in
Table I for a fiducial cosmology $\Theta=(65,0.925,0.3,-0.8,0.3)$
assuming no prior information and
$z_{\rm max}=1.5$.
This particular cosmology was
chosen since, firstly, it is in the middle of the parameter range preferred
by the current data and, secondly, it corresponds to a particular dark energy
model which one might want to constrain~\cite{BM}.
We have tested the stability of our results to
small changes in the parameters compatible with the current
observational data. 

\begin{figure}[!h]
\setlength{\unitlength}{1cm}
\vspace{-0.8cm}
\centerline{\hspace{-0.5cm}\psfig{file=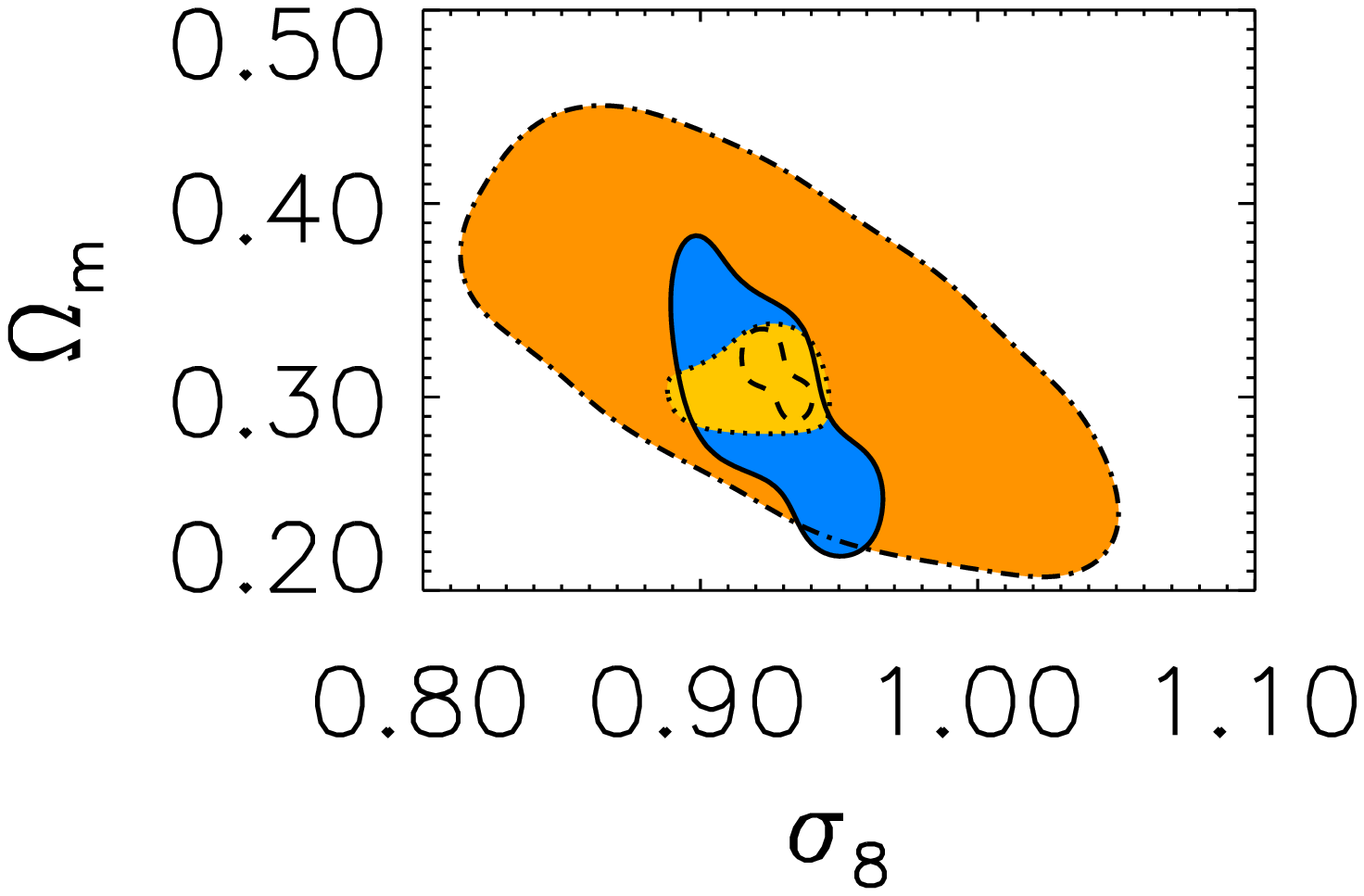,width=5.5cm,height=5.5cm}\hspace{-1.0cm}\psfig{file=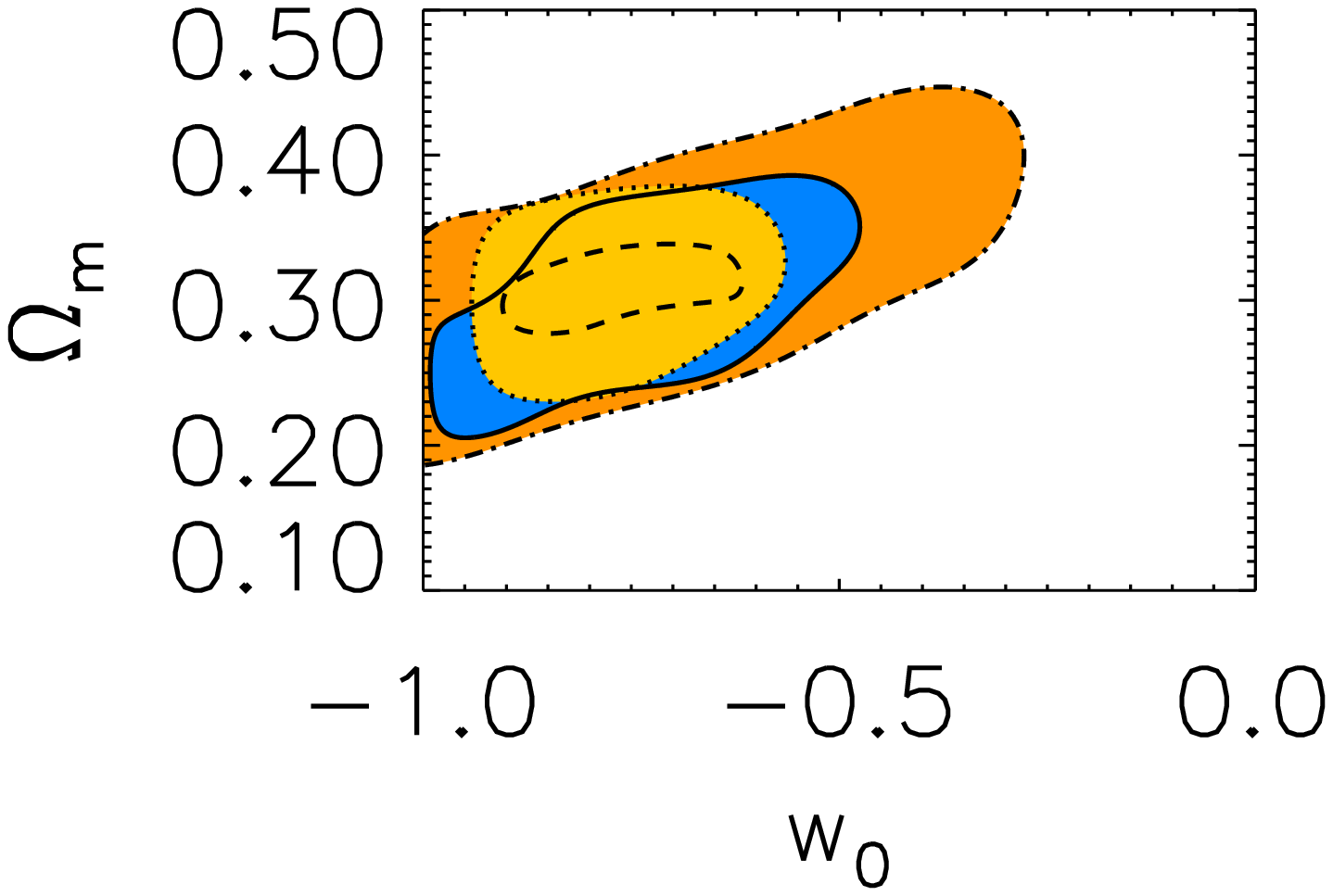,width=5.5cm,height=5.5cm}}
\caption{The marginalized joint likelihood contours in the
$\sigma_8$-$\Omega_{\rm m}$ (left) and $w_0$-$\Omega_{\rm m}$ (right)
planes at the 1-$\sigma$ level. The largest contour corresponds to a
type (I) survey, the dark grey contour to type (II), the light grey contour
to type (III) and the dashed line contour to type (IV).}
\label{fig:cont1}
\end{figure}

The dependence of $dN\over dz$ on $H_0$ is very weak and the double-valued nature of growth factor around $w\approx -0.5$ leads to 
a degeneracy with the amplitude $\sigma_8$. Therefore, it seems sensible to consider the
possibility of prior assumptions on these two parameters, particularly
since both should be measured independently of the properties of the
dark energy by other
means. $H_0$ is  measured using the Hubble Space Telescope at
present to within $\Delta H_0=8$~\cite{Freedman:01}. We will assume that in the next few years a precise
measurement will be possible to $\Delta H_0=5$. In the case of
$\sigma_8$ we will assume that it can be measured almost exactly by, for
example, a low-$z$ X-ray survey. Although this will not be precisely
true it is useful for comparison with ref.~\cite{Haiman:01}. The results of imposing the prior on $\sigma_8$ by itself and combined with that on $H_0$ are also listed in Table I.

There is a clear improvement in one's ability to constrain the parameters in
going from a type (I) to type (IV). From the point of view of the dark
energy the salient parameters are $\Omega_{\rm m}$, $w_0$ and
$w_1$ whose errorbars are often asymmetric due to the complicated
shape of the likelihood surface. Including the prior on $\sigma_8$
appears to be useful in removing degeneracies, whereas the
distribution is very flat in the $H_0$ direction and, therefore,
inclusion of a prior on it has little significant effect.

If one uses no prior information with (I), it is only possible 
to  measure $\sigma_8$ and
$\Omega_{\rm m}$ accurately and place an upper bound on $w_0$.
There is no viable constraint on $w_1$ due to the relatively small number of
clusters that one would detect in such a setup.
If one includes both the priors $|\delta\Omega_{\rm m}|\approx
0.04$ and a weak constraint on $w_0$ is possible, but there is still
little information on $w_1$.

The results of (II) and (III) are qualitatively similar with (III)
improving on (II). With no prior information one can constrain
$\sigma_8$ and $\Omega_{\rm m}$ considerably ($|\delta\sigma_8|$=0.03,
$|\delta\Omega_{\rm
m}|$=0.05 for (II) and $|\delta\sigma_8|$=0.02,
$|\delta\Omega_{\rm m}|$=0.03 for (III)), and
good information on $w_0$ would be possible. However, yet again very
little information would be possible on $w_1$, a situation which
is only mildly alleviated by the inclusion of the prior information. 
It is worth noting that for our chosen
fiducial model it is easier to set an upper bound on $w_1$ than a
lower bound. This is a general trend we observed for the models we
studied, though for some it was less pronounced.

Only in the case of (IV) with a fixed value of $\sigma_8$ can very
strong statements be made about $w_1$ using this kind of
observation. Such a setup also gives very accurate information on all the
other parameters including $w_0$, irrespective any prior. This
provides clear motivation for considering the feasibility of this 
setup.

\begin{figure}[h]
\setlength{\unitlength}{1cm}
\vspace{-0.8cm}
\centerline{\hspace{-0.5cm}\psfig{file=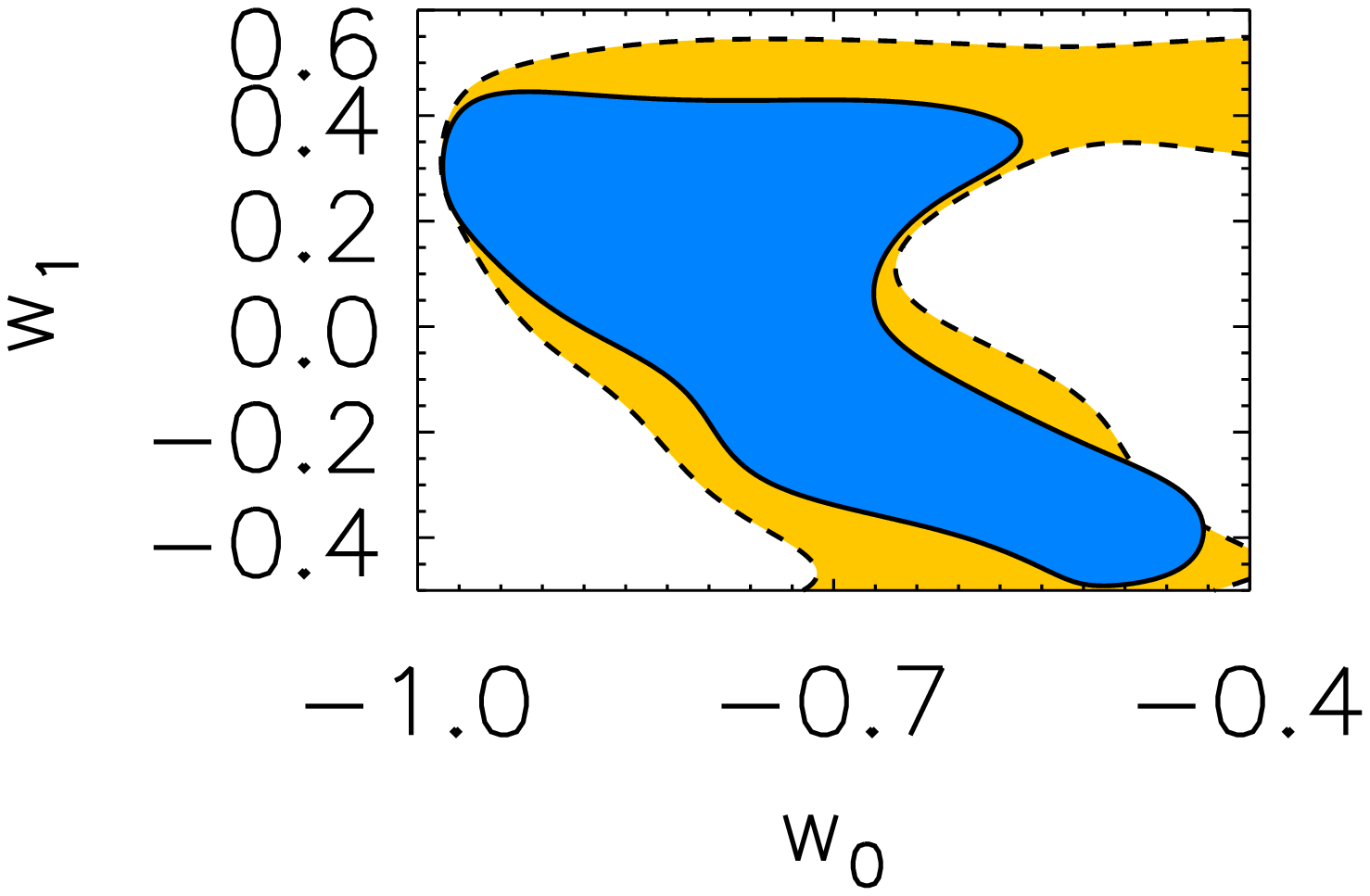,width=5.5cm,height=5.5cm}\hspace{-1cm}\psfig{file=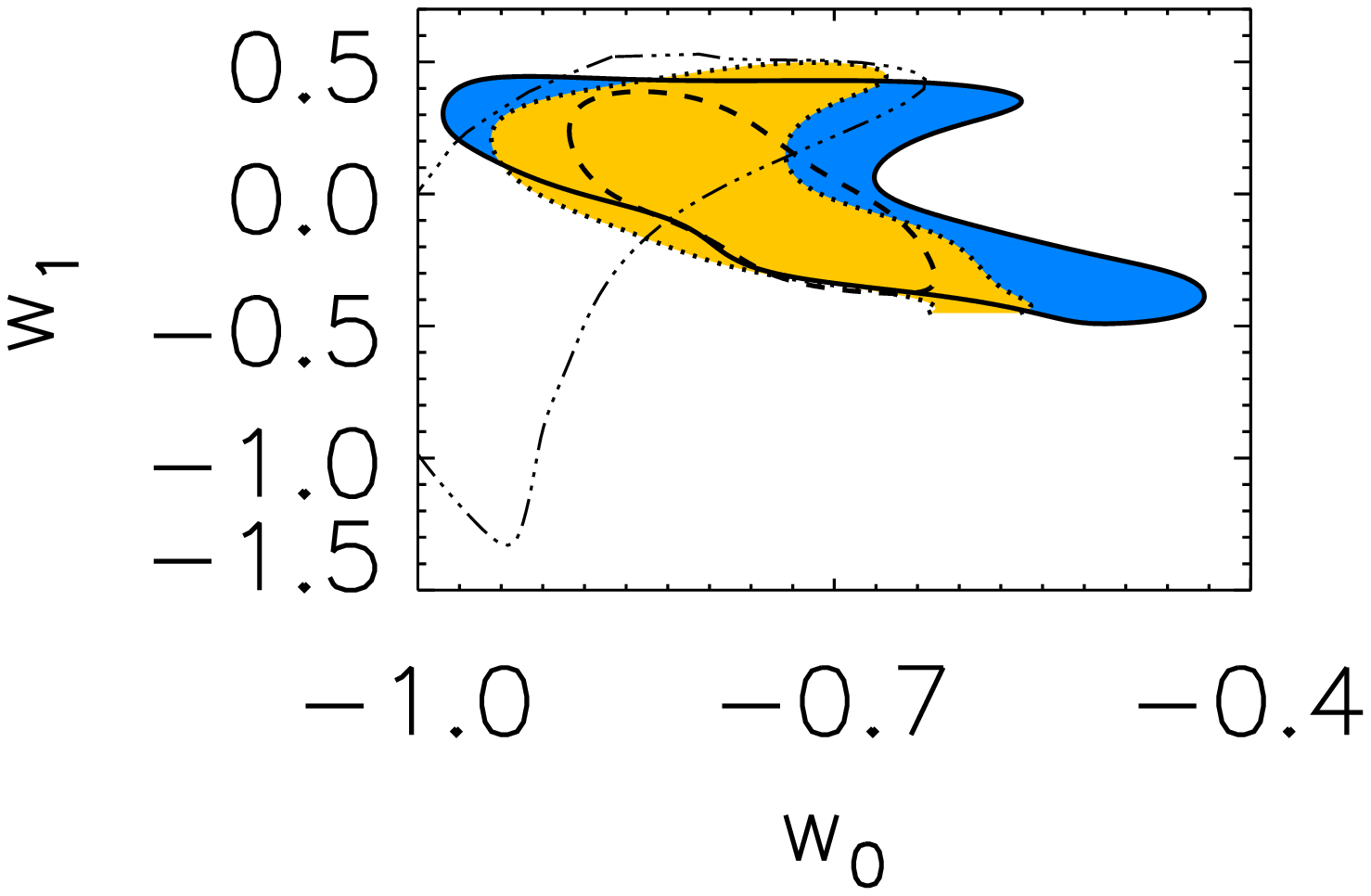,width=5.5cm,height=5.5cm}}
\caption{The 1-$\sigma$ joint likelihoods in the $w_0$-$w_1$ plane. In
the left panel for setup (II) where dark shaded region is obtained with a maximum redshift
of $z_{\rm max}=1.5$ and the light shaded region corresponds to a maximum
redshift of $z_{\rm max}=0.5$. In the right panel for setups
(II), (III) and (IV) using the same conventions as in
Fig.~\ref{fig:cont1}. The transparent 3-dot dashed line corresponds to
the joint likelihood for the SNe survey SNAP.}
\label{fig:cont3}
\end{figure}

We have already noted that the errorbars are in general very
asymmetric. In order to investigate this we have plotted in
Fig.~\ref{fig:cont1} the joint likelihood surfaces in the
$\sigma_8$-$\Omega_{\rm m}$ and $w_0$-$\Omega_{\rm m}$ planes for each of
the setups (I)-(IV), which show visually the relative improvement that
one might expect. The degeneracies are similar to those observed in
previous work~\cite{Haiman:01} and we see that only (II), (III) and
(IV) constrain $w_0$ in any significant way.
Nonetheless, it is clear that in each case the value of $\Omega_{\rm m}$
is constrained extremely well. We have used $z_{\rm max}=1.5$,
however, the using $z_{\rm max}=0.5$ remarkably has little effect on
the size of the errorbars, since it is the statistical weight 
of the large number of clusters found at low redshifts which 
fixes these parameters. We
also performed an analysis with $\Delta z = 0.025$  and
found that this increases the uncertainties on the estimated
parameters in a similar way to changing $z_{\rm max}=1.5$ to
$z_{\rm max}=0.5$. 

The degeneracy between $w_0$ and $w_1$ is particularly important from
the point of view of dark energy and this is illustrated in
Fig.~\ref{fig:cont3}, left panel, for (II) under the optimistic assumption 
that $z_{\rm max}=1.5$ and when $z_{\rm max}=0.5$. The degeneracy has a
complicated, double-valued shape and the constrained region is much
smaller when $z_{\rm max}$ is larger. This is as expected since
constraining these parameters requires more information at high
redshifts. 

Our results show that only for setup (IV) and an effectively
fixed value of $\sigma_8$ can one independently fix the crucial
parameter  $w_1$ using this kind of measurement. However, all is not
lost; it was pointed out in ref.~\cite{AW} that given independent  prior
information on $\Omega_{\rm m}$, SNe measurements can accurately
constrain the dark energy. As we have already pointed out even setup
(I) will provide important information in this respect and the others
will improve on this. 

Even more information can be gleaned by making the comparison of the
two different probes of dark energy in the $w_0$-$w_1$
plane. Fig.~\ref{fig:cont3}, right panel, illustrates this for setups
(II), (III) 
and (IV) compared to a similar calculation for SNAP~\cite{AW}.
Even for (II) performances of the two methods are
comparable in terms of the area of the 1-$\sigma$ contour and for (IV)
the result is very much better. Notice also that the degeneracy in
this plane is also totally different and combining them would give a
localized region pinning down $w_0$ very accurately and $w_1$ to
within $\sim\pm 0.2$. While this may not be enough to rule out $w_1=0$
at the 2-$\sigma$ level, a look at the various models for dark energy
considered in ref.~\cite{AW}, shows that such observations would put tight constraints on the particular dark energy models.

Our basic philosophy has been to investigate the absolute best case
constraints that a given survey can achieve in terms of the properties
of the dark energy. In this spirit, we have shown that as with SNe
observations, cluster surveys selected using the SZ effect will
provide important information as to the nature of the dark energy and
that there is a potential synergy between the two. However, our
conclusions were drawn from a highly idealized model of cluster physics.

One of the key sources of systematic uncertainties will come from the
$M_{\rm lim}-S_{\rm lim}$ relation due to, for example, heat input or the
clusters being not completely virialized~\cite{Verde:01}. These
effects might manifest themselves in terms of either a systematic
shift in the overall normalization, or in statistical scatter. We
have estimated the possible effects of these
uncertainties~\cite{Weller:01b} and found that if  the scatter is less
than about 20\% and the overall normalization is accurate to with 5\%
then one can distinguish our fiducial model from the standard
$\Lambda$CDM model. To get an idea of why the constraint on required
accuracy of the overall normalization is particularly important we
just comment that a 20\% change would lead to a factor of 2 change in
the total number of clusters. 

We are optimistic that many of the practical difficulties which we
have ignored will be addressed with  the first generation of SZ survey
instruments and can be taken into account in the future with the
qualitative picture of our results will remain: that SZ cluster
surveys provide a robust complementary probe for dark energy.

\noindent  ACKNOWLEDGMENTS: We thank R.~Crittenden, M.~Jones and
P.~Wilkinson for useful discussions. The parallel computations were
done at the UK National Cosmology Supercomputer Center funded by
PPARC, HEFCE and Silicon Graphics / Cray Research. JW and RB  are
supported by a PPARC.

\def\jnl#1#2#3#4#5#6{\hang{#1, {\it #4\/} {\bf #5}, #6 (#2).}}
\def\jnltwo#1#2#3#4#5#6#7#8{\hang{#1, {\it #4\/} {\bf #5}, #6; {\it
ibid} {\bf #7} #8 (#2).}} 
\def\prep#1#2#3#4{\hang{#1, #4.}} 
\def\proc#1#2#3#4#5#6{{#1, in {\it #3 (#4)\/}, edited by #5,\ (#6).}}
\def\book#1#2#3#4{\hang{#1, {\it #3\/} (#4, #2).}}
\def\jnlerr#1#2#3#4#5#6#7#8{\hang{#1 (#2), {\it #4\/} {\bf #5}, #6. {Erratum:} {\it #4\/} {\bf #7}, #8.}}
\def\prl{Phys.\ Rev.\ Lett.}
\def\pr{Phys.\ Rev.}
\def\pl{Phys.\ Lett.}
\def\np{Nucl.\ Phys.}
\def\prp{Phys.\ Rep.}
\def\rmp{Rev.\ Mod.\ Phys.}
\def\cmp{Comm.\ Math.\ Phys.}
\def\mpl{Mod.\ Phys.\ Lett.}
\def\apj{Ap.\ J.}
\def\apjl{Ap.\ J.\ Lett.}
\def\aap{Astron.\ Ap.}
\def\cqg{Class.\ Quant.\ Grav.} 
\def\grg{Gen.\ Rel.\ Grav.}
\def\mn{MNRAS}
\def\ptp{Prog.\ Theor.\ Phys.}
\def\jetp{Sov.\ Phys.\ JETP}
\def\jetpl{JETP Lett.}
\def\jmp{J.\ Math.\ Phys.}
\def\zpc{Z.\ Phys.\ C}
\def\cupress{Cambridge University Press}
\def\pup{Princeton University Press}
\def\wss{World Scientific, Singapore}
\def\oup{Oxford University Press}
\def\asj{Astron.~J}
\def\imp{Int.\ J.\ Mod.\ Phys.}

\end{document}